\documentclass[twocolumn,prl,10pt,amsmath,amssymb,nofootinbib,showpacs,superscriptaddress,floatfix]{revtex4-1}

\newcommand{\rs}{\rm\scriptscriptstyle}
\usepackage{graphicx}
\usepackage{color}
\usepackage[usenames,dvipsnames]{xcolor}
\usepackage[colorlinks=true,linkcolor=Red,citecolor=Green,linktoc=page]{hyperref}
\usepackage{multirow}
\usepackage{float}
\usepackage{flushend}
\usepackage{balance}
\usepackage[varg]{txfonts}
\usepackage{ulem}
\usepackage{fancyhdr}

\begin{document}
\title{Charge Berezinskii-Kosterlitz-Thouless transition in superconducting NbTiN films}
\author{Alexey\,Yu.\,Mironov}
\affiliation{A.\,V.\,Rzhanov Institute of Semiconductor Physics SB RAS, 13 Lavrentjev Avenue, Novosibirsk 630090, Russia}
\affiliation{Novosibirsk State University, Pirogova str. 2, Novosibirsk 630090, Russia}
\affiliation{The James Franck Institute and Department of Physics, The University of Chicago, Chicago, IL 60637, USA}
\author{Daniel\,M.\,Silevitch}
\affiliation{Division of Physics, Mathematics, and Astronomy, California Institute of Technology, Pasadena, CA 91125, USA}
\author{Thomas\,Proslier}
\affiliation{Institut de recherches sur les lois fundamentales de l'univers, Commissariat de l'\'{e}nergie atomique et aux \'{e}nergies renouvelables-Saclay, Gif-sur-Yvette, France}
\author{Svetlana\,V.\,Postolova}
\affiliation{A.\,V.\,Rzhanov Institute of Semiconductor Physics SB RAS, 13 Lavrentjev Avenue, Novosibirsk 630090, Russia}
\affiliation{Novosibirsk State University, Pirogova str. 2, Novosibirsk 630090, Russia}
\author{Maria\,V.\,Burdastyh}
\affiliation{A.\,V.\,Rzhanov Institute of Semiconductor Physics SB RAS, 13 Lavrentjev Avenue, Novosibirsk 630090, Russia}
\affiliation{Novosibirsk State University, Pirogova str. 2, Novosibirsk 630090, Russia}
\author{Anton\,K.\,Gutakovskii}
\affiliation{A.\,V.\,Rzhanov Institute of Semiconductor Physics SB RAS, 13 Lavrentjev Avenue, Novosibirsk 630090, Russia}
\affiliation{Novosibirsk State University, Pirogova str. 2, Novosibirsk 630090, Russia}
\author{Thomas\,F.\,Rosenbaum}\affiliation{Division of Physics, Mathematics, and Astronomy, California Institute of Technology, Pasadena, CA 91125, USA}
\author{Valerii\,M.\,Vinokur}
\affiliation{Materials Science Division, Argonne National Laboratory, 9700 S. Cass Ave, Argonne, IL 60439, USA}
\author{Tatyana\,I.\,Baturina}
\affiliation{A.\,V.\,Rzhanov Institute of Semiconductor Physics SB RAS, 13 Lavrentjev Avenue, Novosibirsk 630090, Russia}
\affiliation{Novosibirsk State University, Pirogova str. 2, Novosibirsk 630090, Russia}
\affiliation{The James Franck Institute and Department of Physics, The University of Chicago, Chicago, IL 60637, USA}
\affiliation{Departamento de F\'isica de la Materia Condensada, Instituto de Ciencia de Materiales Nicol\'as Cabrera and Condensed Matter Physics Center (IFIMAC), Universidad Aut\'onoma de Madrid, 28049 Madrid, Spain}
\begin{abstract}
A half-century after the discovery of the superconductor-insulator transition (SIT), one of the fundamental predictions of the theory, the charge Berezinskii-Kosterlitz-Thouless (BKT) transition that is expected to occur at the insulating side of the SIT, has remained unobserved. 
The charge BKT transition is a phenomenon dual to the vortex BKT transition, which is at the heart of the very existence of two-dimensional superconductivity as a zero-resistance state appearing at finite temperatures. The dual picture points to the possibility of the existence of a superinsulating state endowed with zero conductance at finite temperature.  
 Here, we report the observation of the charge BKT transition on the insulating side of the SIT, identified by the critical behavior of the resistance.
 We find that the critical temperature of the charge BKT transition depends on the magnetic field exhibiting first the fast growth and then passing through the maximum at fields much less than the upper critical field.  Finally, we ascertain the effects of the finite electrostatic screening length and its divergence at the magnetic field-tuned approach to the superconductor-insulator transition. 
\end{abstract}

\maketitle

\newpage
In 1996 Diamantini \textit{et al}.\,\cite{Diam:1996} demonstrated that in planar Josephson junction arrays (JJA) the vortex-charge duality leads to a zero-temperature quantum phase transition between a superconductor and its mirror image, which they termed a superinsulator. The physical origin of a superinsulating state is the charge confinement due to the logarithmic interaction between the charges in two-dimensional (2D) systems\,\cite{Fazio:1991,nature,BVAnnals:2013}. In disordered superconducting films, the charge confinement on the insulating side of the SIT results from the divergence of the dielectric constant $\varepsilon$ in the critical vicinity of the transition. The logarithmic interaction holds over distances $d<r\lesssim\Lambda\simeq\varepsilon d$, where $d$ is the thickness of the film and $\Lambda$ is the electrostatic screening length\,\cite{BVAnnals:2013}. This parallels the logarithmic interaction between vortices on the superconducting side, which causes the vortex binding-unbinding topological BKT transition into the superconducting state at finite temperature $T=T_{\rs VBKT}$\,\cite{Ber,KT1972,KostThoul}. Accordingly, logarithmic interactions between charges on the insulating side of the SIT is expected to give rise to a charge BKT transition into the superinsulating state with the conductance going to zero at a finite temperature $T=T_{\rs CBKT}$\,\cite{nature,BVAnnals:2013,Anderson:1979}. 
Applied magnetic fields can tune the SIT with high resolution, offering a window into unexplored electronic functionalities since in the critical vicinity of the SIT at the superconducting side the system should possess the superinductance\,\cite{Ioffe2012}, and at the insulating side the system is expected to be a supercapacitor due to diverging dielectric constant\,\cite{BVAnnals:2013}. This calls for a thorough study of the highly resistive state that terminates two-dimensional superconductivity at the quantum critical point whose nature remains a subject of intense research\,\cite{Zvi:1994,Samba:2005,Kapitulnik:2005,BMV:2007,nature,BVHyperact:2008,Shahar:2015srep}.

Existing experimental data on JJA\,\cite{Kanda:1995,Kanda:1996}, superconducting wire networks\,\cite{Baturina:2011}, InO\,\cite{Shahar:2015srep}, and TiN films\,\cite{BVHyperact:2008,Kalok:2012} support the picture of the dual vortex-charge BKT transitions and corresponding formation of the mirror superconducting-superinsulating states.
Yet, while there have been numerous experimental hallmarks of superinsulating behavior, the evidence for the charge BKT transition, with its characteristic criticality, has remained elusive. To answer this challenge, we examine a NbTiN film,
which is expected to combine the high stability of TiN films with the enhanced superconducting transition temperature $T_c$ of NbN films, due to a larger Cooper pairing coupling constant as compared to TiN.
We thus expect that all other relevant temperature scales, including the $T_{\rs CBKT}$, are enhanced as well, 
opening a wider window for observing critical behavior.

\begin{figure*}[t!]
\centering
\includegraphics[width=.9\linewidth]{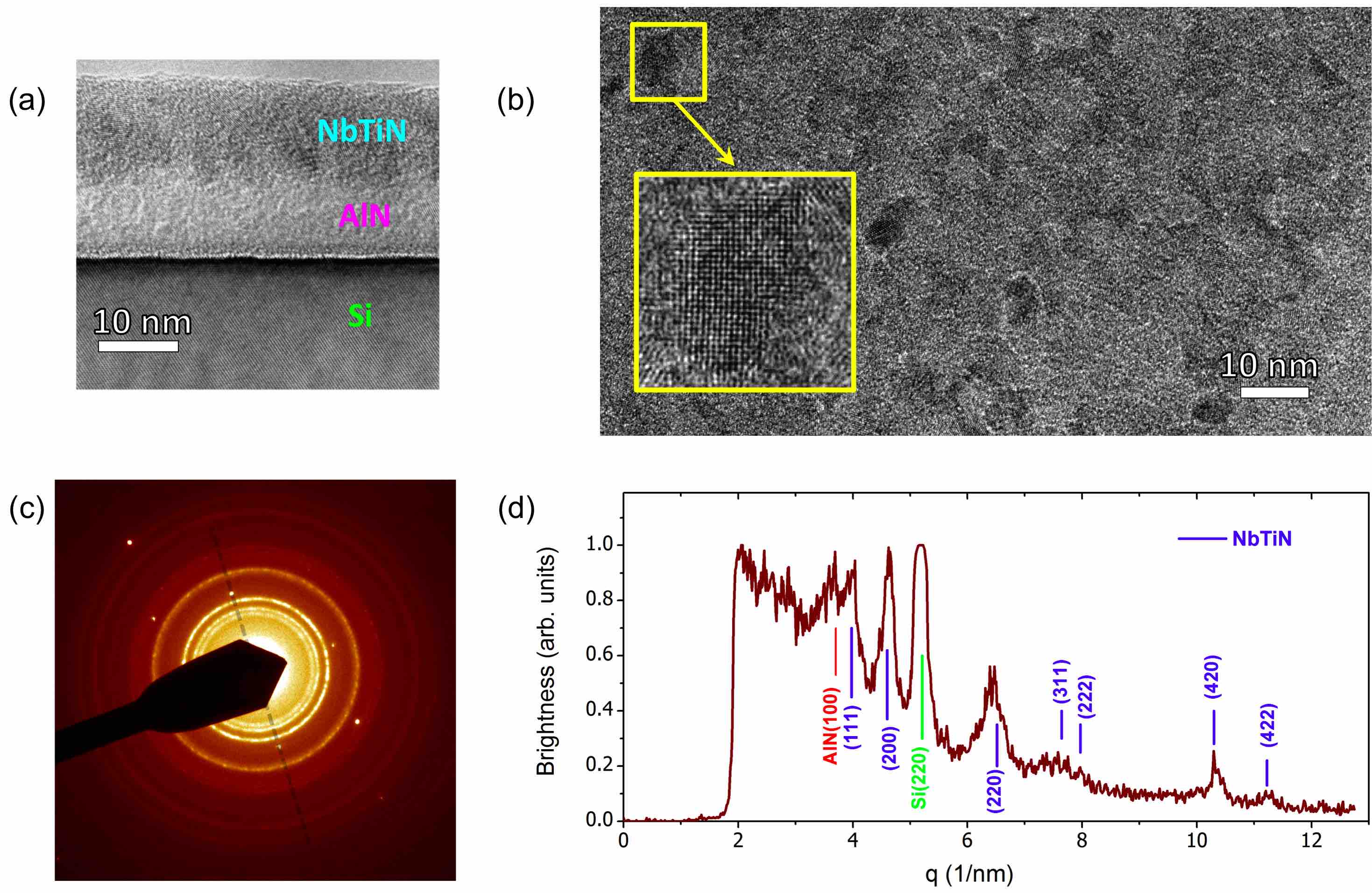}
\caption{ Structure and composition of a 10 nm thick NbTiN film.
(a) Cross-section of the film from High Resolution Transmission Electron Microscopy (HRTEM).
(b) HRTEM plan view bright field image. The yellow square shows a magnified image of one of the crystallites.
(c) Electron-diffraction data of the film.
The rings are characteristic of polycrystalline structures; the bright spots arise from the crystalline lattice of the underlying Si substrate.
(d) Electron diffraction data from the panel (c) taken along the self-transparent dashed line in panel (c) and plotted as the intensity vs. the wavenumber. Several Bragg peaks from the NbTiN film are observed, along with the (220) peak from the Si substrate and the (100) peak from the AlN buffer layer.}
    \label{fig:fig1}
\end{figure*}

To grow suitable NbTiN films, we employed the atomic layer deposition (ALD) technique based on sequential surface reaction step-by-step film growth. The fabrication technique is described in Supplementary Materials (SM). This highly controllable process provides superior thickness and stoichiometric uniformity and an atomically smooth surface\,\cite{Lim:2003}. We used NbCl$_5$, TiCl$_4$, and NH$_3$ as gaseous reactants; the stoichiometry was tuned by varying the ratio of TiCl$_4$/NbCl$_5$ cycles during growth\,\cite{Proslier:2011}. The superconducting properties of these ultrathin NbTiN films were optimized by utilizing AlN buffer layers grown on top of the Si substrate\,\cite{Shiino:2010}.
NbTiN films of thicknesses $d=10$,\,15, and 20\,nm were grown, varying only the number of ALD cycles (240, 420, and 768 cycles, respectively), with all other parameters of the ALD process held constant. 
We show in Fig.\,\ref{fig:fig1}(a) a high-resolution transmission electron microscopy (HRTEM) image of the cross-section of the 10 nm thick NbTiN film. 
It reveals that both the AlN buffer layer and the NbTiN have a fine-dispersed polycrystalline structure. 
Presented in Fig.\,\ref{fig:fig1}(b) is a plan view of a  large area containing many crystallites. The densely packed crystallites have different orientations and are separated by atomically thin inter-crystallite boundaries. A statistical analysis of the image finds the average crystallite size to be approximately 5\,nm. The electron diffraction data for the film are shown in Fig.\,\ref{fig:fig1}(c). The clearly seen rings confirm a polycrystalline structure.
The analysis of the diffraction data along the direction [220] of the Si substrate displayed in Fig.\,\ref{fig:fig1}(d) reveals that the NbTiN crystallites have the same rock-salt crystal structure as both NbN and TiN. Using Vegard's law, we find that our NbTiN film is an approximately 7:3 solid solution of NbN and TiN (see SM).

\begin{figure*}[t!]
\centering
		\includegraphics[ width=0.85\linewidth]{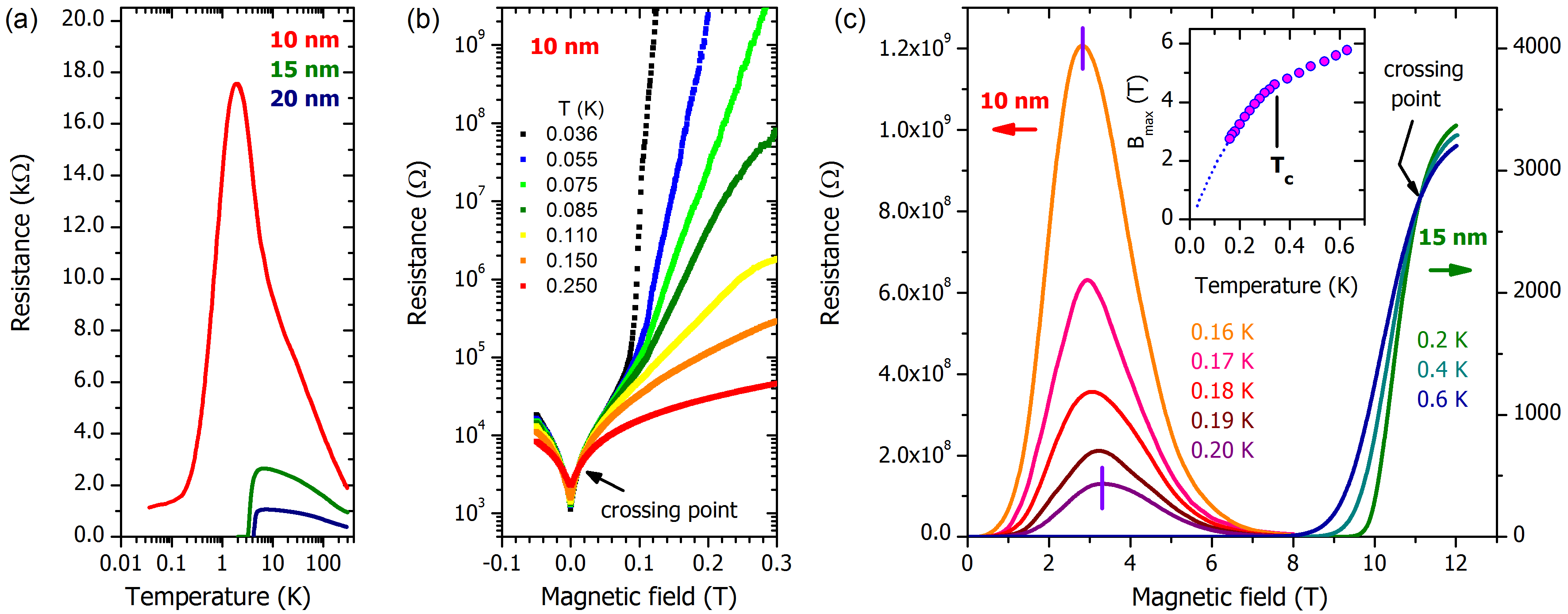}
	\caption{
Temperature and magnetic field dependences of the resistance of NbTiN films.
(a) The temperature dependences of the resistance in zero magnetic field for three NbTiN films of thicknesses 10, 15, and 20\,nm respectively.
(b) Low-field isothermal magnetoresistance of the 10\,nm thick film at different temperatures.
At temperatures below 0.1\,K, the magnetoresistance develops a sharp kink at $B<$0.1\,T having the trend of moving to lower magnetic fields upon decreasing temperature. Above the kink, even a small increase in the field results in a sharp increase of the resistance by several orders of magnitude. The crossing point, $R_c=4.7$\,k$\Omega$, $B_{\rs SIT}=0.015$\,T, separates the regions with $dR/dT<0$ and $dR/dT>0$.
(c)  Linear magnetoresistance of the 10\,nm and 15\,nm thick films. 
The left ordinate corresponds to the data for 10\,nm thick film taken in the 0.16 -- 0.20\,K temperature range, and the right hand ordinate refers to the data in the 0.2 -- 0.6\,K interval for the 15\,nm thick film.
Note that the two resistance scales differ by a factor of $3\cdot 10^5$.
The magnetoresistance of the 15\,nm thick film exhibits the conventional superconducting behavior with the well-defined upper critical field $B_{c2}(0)=$10.5\,T and the crossing point at $B_c=11$\,T stemming from the interplay of superconducting fluctuations contributing to conductivity\,\cite{Baturina2005,Finn}.  By contrast, the magnetoresistance of the 10\,nm thick film develops a colossal insulating peak at fields well below $B_{c2}$.
The vertical strokes on the 0.16\,K and 0.20\,K curves for the 10\,nm thick film mark the fields $B_{\mathrm{max}}$ at which the magnetoresistance peaks are achieved. 
The inset presents the temperature dependence of $B_{\mathrm{max}}$ (symbols). The dotted line, extrapolating the data to $T\to 0$ limit, illustrates the trend of $B_{\mathrm{max}}$  of shifting towards almost zero field upon decreasing temperature.
 	}
	\label{fig:fig2}
\end{figure*}

The films were lithographically patterned into bars and resistivity measurements were performed at sub-Kelvin temperatures in helium dilution refrigerators (see the details of the sample geometry and measurement technique in SM). Upon cooling in zero magnetic field, all three films undergo a superconducting transition. 
The temperature dependences of the resistance, $R(T)$, given as resistance per square, are shown in Fig.\,\ref{fig:fig2}(a) over four decades in temperature. 
All the data presented in this paper were measured in the linear response regime.  
The superconducting transition temperature, $T_c$, decreases with decreasing film thickness and consequent increasing sheet resistance. The resistances of all three films exhibit peaks at temperatures just above $T_c$, with the peak amplitudes increasing as the thickness decreases.  Similar trends were observed in the parent compounds TiN\,\cite{Baturina:2012} and NbN\,\cite{Pratap:2012} near the SIT and were attributed to quantum contributions to conductivity due to weak localization and electron-electron interaction effects. 
The sheet resistance of the thinnest film achieves a maximum of 17.56\,k$\Omega/\square$; notably, this well exceeds the quantum resistance $R_Q=h/(2e)^2=6.45\,$k$\Omega/\square$ which is widely believed to be the upper boundary for the existence of superconductivity in two dimensions. 
A similar peak of 29.4\,k$\Omega/\square$, well above $R_Q$, was seen in TiN\,\cite{Baturina:2007}.

Focusing on the behavior of the thinnest film ($d = 10$~nm), we note first that the global coherent superconducting state is not achieved at lowest temperatures.
Instead, the behavior of the zero field resistivity suggests that the
film falls into the Bose metallic state, featuring a finite  density of free vortices.
Figure\,\ref{fig:fig2}(b) presents a set of magnetoresistance curves, $R(B)$, taken at different temperatures below $T_c=0.33$\,K determined by the inflection point of $R(T)$ at zero magnetic field. Prominent features of these magnetoresistance curves, especially profound at lowest temperatures, are the crossing point at very low field $B_{\rs SIT}=0.015$\,T that marks the SIT and will be discussed later and the sharp kink at some temperature-dependent magnetic field above which the resistance increases extremely quickly as a function of the field. 
The kink field shifts towards lower fields upon decreasing temperature.
Inspecting the magnetoresistance behavior in the large field interval, one sees that it shows inherently non-monotonic behavior marked by a colossal insulating peak, see Fig.\,\ref{fig:fig2}(c).
Importantly, these peaks develop at magnetic fields for which the thicker films are still fully superconducting, i.e. the field $B_{\mathrm{max}}$, where the maximum is observed, is well below the upper critical field $B_{c2}$.
This suggests that the indefinite growth of $R(B)$ at low temperatures/magnetic fields as well as the peak in the resistance at higher temperatures in the 10\,nm film is an implication of Cooper pairing.
The inset in Fig.\,\ref{fig:fig2}(c) shows that the position of the maximum of the resistance peaks moves to lower fields  upon decreasing temperature.
There is a kink in the $B_{max}(T)$ dependence at $T\simeq T_c$, with the slope decreasing signficantly when passing to $T>T_c$. 
Extrapolating the data to $T\to 0$ shows that $B_{max}(T)$ shifts to nearly zero field upon decreasing the temperature.
Taken together, this indicates that the mechanism that drives the system into the strongly localized state overpowers the effect of the suppression of the Cooper pairing by the magnetic field, which would be expected to  diminish the resistance of the Cooper pair insulator.

\begin{figure*}[t!]
	\centering
	\includegraphics[width=0.85\linewidth]{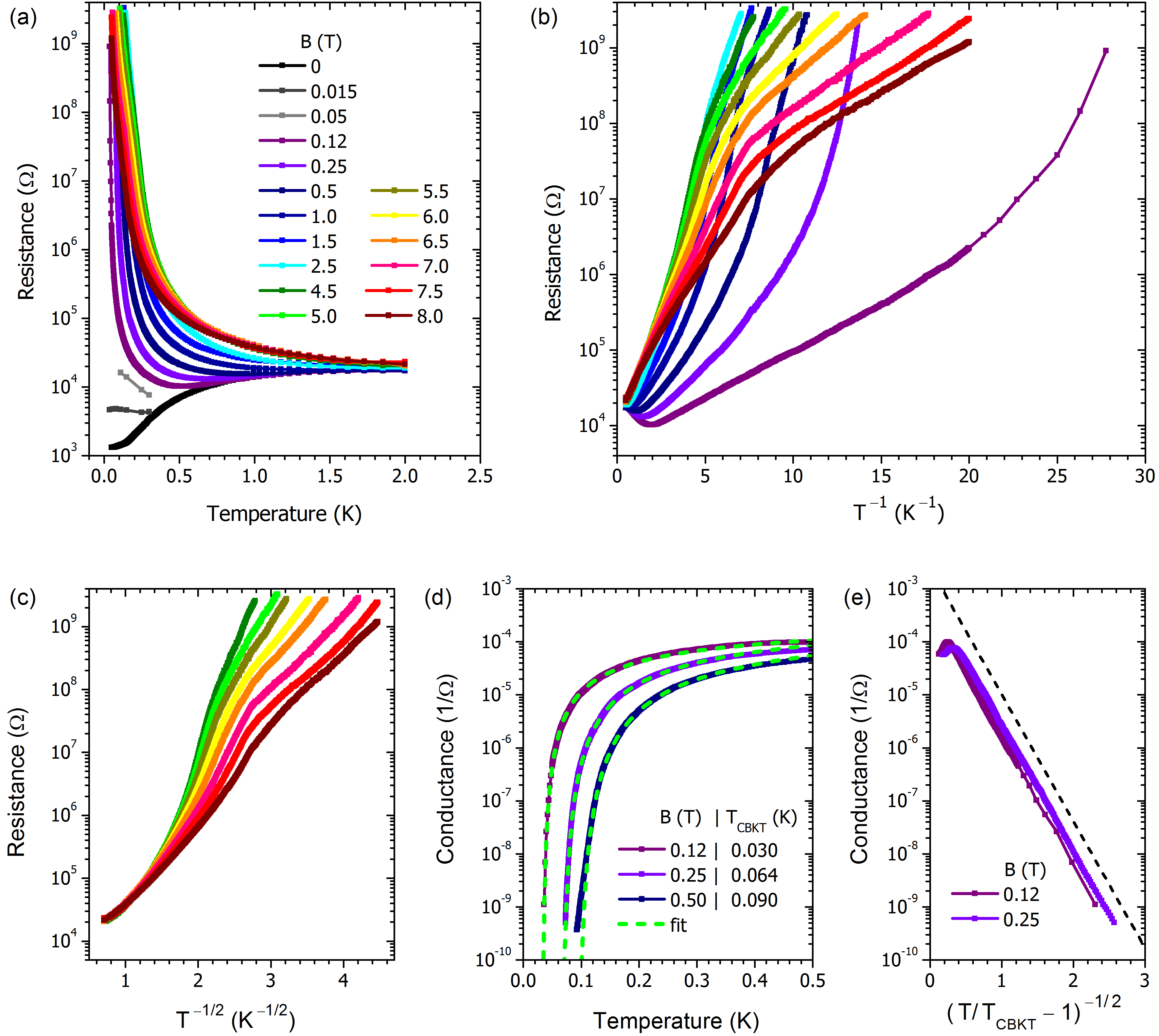}
	\caption{
Temperature evolution of the magnetic field-induced states in the 10 nm thick  NbTiN film.
(a) Resistance in the log scale vs. temperature at different magnetic fields listed in the legend in panel. 
The legend is the same for all the panels of this figure.
(b) Resistance vs. inverse temperature at different magnetic fields. 
Since none of the traces is a straight line, none of the temperature dependences can be viewed as an Arrhenius behavior with a unique activation energy across the entire temperature interval.
(c) Resistance in the log scale vs. $T^{-1/2}$ in the field range between 4.5 and 8\,T.
(d) Three representative curves from panel (a) replotted as a conductance $G = 1/R$, vs. temperature, demonstrating the transition into a superinsulating state at finite temperature.
The dashed lines show the best fits to Eq.\,(\ref{eq:ChargeBKT}) with the corresponding $T_{\rs CBKT}$ listed in the legend.
(e) The conductance as a function of $(T/T_{\rs CBKT}-1)^{-1/2}$ for two magnetic fields.
The dashed black line is a guide to the eye revealing that linear slopes of the two curves are the same.
	}
	\label{fig:RTBi}
\end{figure*}

To gain insight into the nature of the magnetic field-induced states, we examine $R(T)$ at different magnetic fields. Figure\,\ref{fig:RTBi}(a) displays the fan-like set of magnetoresistance
curves, characteristic to the magnetic field-induced SIT, for the 10\,nm thick film. 
The crossing point $(B_{\rs SIT},R_c)$ in Fig.\,\ref{fig:fig2}(b) now corresponds to nearly temperature independent $R(T)$ at $B=0.015$\,T separating between the superconducting and insulating behaviors. 
Two important comments are in order here. First, the field of the crossing point $B_{\rs SIT}=0.015$\,T is \textit{by two orders of magnitude lower} than the upper critical field $B_{c2}$. 
This differs it from the crossing point displayed by the thicker film ($d=15$\,nm) occurring at 11\,T and resulting from the quantum contributions to conductivity from superconducting fluctuations\,\cite{Baturina2005,Finn}. 
Second, the resistance at the SIT is $R_c=4.7$\,k$\Omega$ that is close but not equal to quantum resistance 6.45\,k$\Omega$.

Re-plotting these data as log$R$ vs.\,$1/T$ curves in Fig.\,\ref{fig:RTBi}(b),
we see that the behavior of the resistance in the entire temperature range cannot be reduced to the Arrhenius temperature dependence with a single activation energy. 
The activation mechanism would have manifested itself as a straight line on this graph.
Instead there is a complicated evolution of the resistance curves with increasing magnetic field. 
While at low fields the log$R(1/T)$ dependences exhibit hyperactivation, i.e. faster than thermally activated growth\,\cite{BVHyperact:2008}, at larger fields, the log$R(1/T)$ curves exhibit a kink and bend down with decreasing temperature. 
Note that in contrast to what was shown for InO\,\cite{Shahar:2015srep}, our bending down curves are inconsistent with the Efros-Shklovskii behavior, see Fig.\,\ref{fig:RTBi}(c).

In order to illuminate the physics governing the $R(T)$ behavior, we replot the low-field data as the conductances, $G = 1/R$, vs. temperature in Fig.\,\ref{fig:RTBi}(d). 
We see in the conductance curves an insulating analogue of the drop to zero of the resistance at the onset of superconductivity. 
In the dual mirror picture of the conductance, we thus see the transition of the system into a superinsulating state characterized by zero conductance at finite temperature. 
This suggests that we can write the conductance in the generic form $\ln G\propto -a/(T-T^*)^{\alpha}$ for the finite temperature zero conducting state. 
Using $T^*$ as an adjusting parameter, we find that $\alpha = 0.48 \pm 0.03$ gives the best fit to the experimental data, consistent with $\alpha=1/2$ corresponding to critical BKT behavior. We thus arrive at the following expression for conductivity: 
    \begin{equation}
        G=G_0\exp\left(-\frac{b}{\sqrt{(T/T_{\scriptscriptstyle{\mathrm{CBKT}}})-1}} \right)\,,
        \label{eq:ChargeBKT}
    \end{equation}
 with $T_{\rs CBKT}$ replacing $T^*$. 
 In Fig.\,\ref{fig:RTBi}(e) we plot $G$ vs. $(T/T_{\rs CBKT}-1)^{-1/2}$ for fields 0.12 and 0.25\,T. The correct choice of the only adjustable parameter, $T_{\rs CBKT}$, for each field (shown in the legend for Fig.\,\ref{fig:RTBi}(d)), produces a linear dependence allowing the determination of $b$ and $G_0$ as the slope and the intercept of the respective lines:
$b$ was found to be 5.5 independent of magnetic field, whereas $G_0$ varied with field.
The dashed lines in  Fig.\,\ref{fig:RTBi}(d) correspond to Eq.\,(\ref{eq:ChargeBKT}). 
A close inspection of the fits reveals that while the conductance curves at $B=0.12$ and 0.25\,T precisely follow the formula over six decades, the 0.50\,T curve displays a slight departure from the critical CBKT dependence. 

\begin{figure}[t!]
	\centering
	\includegraphics[ width=1\linewidth]{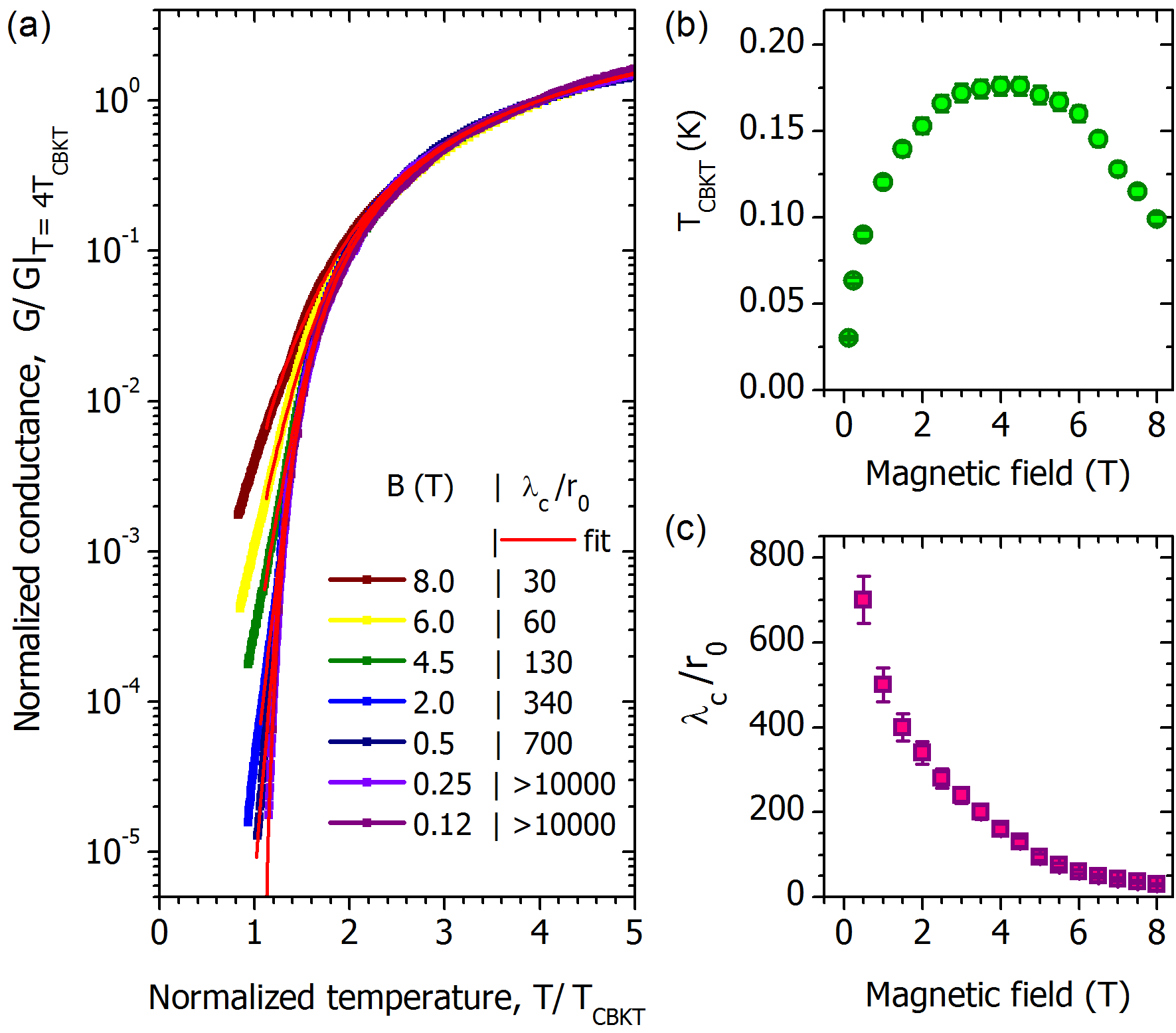}
	\caption{Charge BKT in the 10 nm NbTiN film. 
		(a) Normalized conductance vs. normalized temperature. 
		The temperature is normalized with respect to $T_{\rs CBKT}$ and the conductances were normalized by their values at high temperatures $T=4T_{\rs CBKT}$.
		Symbols stand for experimental data, and red lines show the
		fit obtained from the self-consistent solution to Eqs.\,(\ref{eq:nf}),(\ref{eq:Eq}). We present data covering the full range of magnetic field but omit a few curves to avoid overcrowding the plot. 
		(b) Magnetic field dependence of the transition temperatures $T_{\rs CBKT}$.
		(c) The screening length $\lambda_c$ in the units of $r_0$ vs. magnetic field. 
	}
	\label{fig:fig4}
\end{figure}
 
To proceed further with the analysis, we choose the value of $T_{\rs CBKT}$ for every isomagnetic $G(T)$ curve and plot in Fig.\,\ref{fig:fig4}(a) the conductance normalized by its value at temperature $T=4T_{\rs CBKT}$ vs. the normalized temperature $T/T_{\rs CBKT}(B)$. 
The corresponding field-dependent charge BKT transition temperature $T_{\rs CBKT}$ is shown in Fig.\,\ref{fig:fig4}(b). 
Remarkably, the complex diversity of the $R(T)$ Arrhenius plots of Fig.\,\ref{fig:RTBi}(b), collapse onto the universal curve. Moreover, the field-dependent evolution of the curve shapes, including the change in concavity, now reduces simply to a successive 
deviation from the universal curve: the higher the magnetic field at which the curve is measured, the higher the $T/T_{\rs CBKT}$ ratio at which the given curve departs from the universal envelope.
We stress that the above normalization procedure does not presume any special temperature dependence of $G(T)$. The choice of the temperature at which the normalizing value of the conductance is taken is somewhat arbitrary, as seen from the quality of the collapse  over the {2.5 $\div$ 5 range in $T/T_{\rs CBKT}$.
 
 We now describe the overall $G(T,B)$ behavior using the two dimensional Coulomb gas model developed in Ref.\,\cite{Minnhagen:1987}.
We note first that the conduction is controlled by the density of free charge carriers, $G\propto n_f$, i.e. the conductance is proportional  to the inverse squared mean distance between the carriers. In the critical BKT region, $n_f$ is the 2D density of the unbound charges, which is related to the correlation length $\lambda$ at which the unbound charges appear via the equation:
\begin{equation}
n_f=\frac{1}{2\pi}\bigg(\frac{1}{\lambda^2}-\frac{1}{\lambda_c^2}\bigg) \,,
\label{eq:nf}
\end{equation}
where $\lambda_c$ is the smaller of  the bare electrostatic screening length of the film, $\Lambda$ or the lateral linear dimension of the film.
The screening length defines the maximal spatial scale of the logarithmic charge interaction in the film $V(r)\propto\ln(r/\Lambda)$ for $r_0< r<\Lambda$. 
The short distance cutoff, $r_0$, is of the order of the film thickness.
Relating $\lambda_c$ and the density of the unbound charges through the Poisson equation, we derive, following\,\cite{Minnhagen:1987}, a self-consistent equation for $\lambda$
\begin{equation}
\bigg(\frac{\lambda}{r_0}\bigg)^{\sqrt{t-1}}
={\cal Z}\bigg(1-\frac{\lambda^2}{\lambda_c^2}\bigg) \,,
\label{eq:Eq}
\end{equation}
where $t=T/T_{\rs CBKT}$, and ${\cal Z}$ is the constant.
At $\lambda_c\to\infty$, Eqs.\,(\ref{eq:nf}),(\ref{eq:Eq}), reduces to Eq.\,(\ref{eq:ChargeBKT}), with $b=2\ln(1/{\cal Z})$.

Solving Eqs.\,(\ref{eq:nf}) and (\ref{eq:Eq}) with $\lambda_c$ the only adjustable parameter, we find an excellent fit to all measured isomagnetic $G(T)$ (see representative curves in Fig.\,\ref{fig:fig4}(a)). The field dependence of $\lambda_c$ is given in Fig.\,\ref{fig:fig4}(c). Thus, equations (\ref{eq:nf}) and (\ref{eq:Eq}) of the two-dimensional Coulomb gas model completely describe all the complex diversity of the experimental data, including both the BKT critical behavior and the deviation from criticality, using only one adjustable parameter. This establishes a superinsulator as a confined low temperature charge BKT phase of the Cooper pair insulator. 
In this phase, vortices form a Bose condensate that completely blocks the motion of the Cooper pairs.

\begin{figure}[t!]
	\centering
	\includegraphics[ width=1\linewidth]{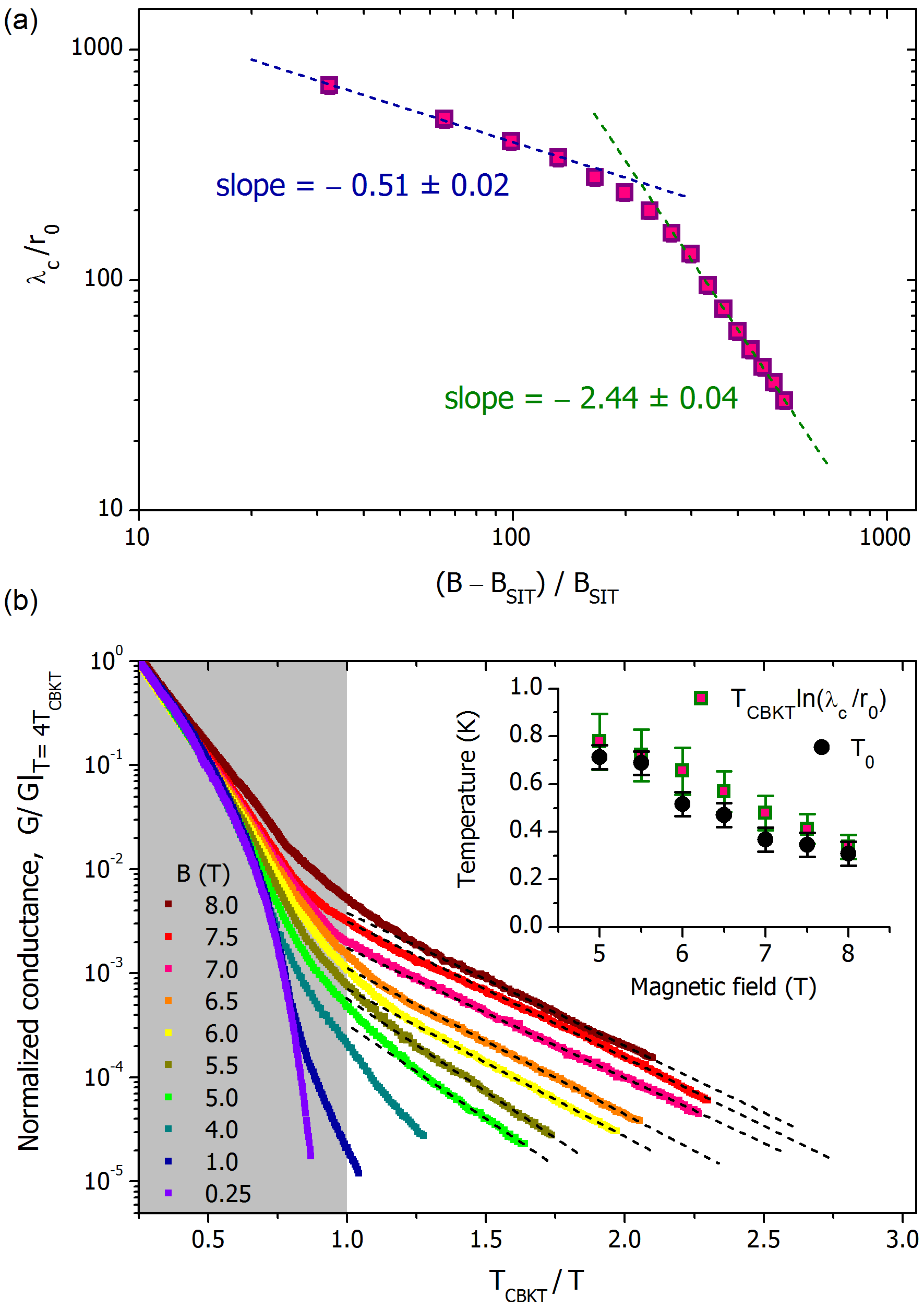} 
	\caption{Critical behavior and the effects of the finite electrostatic screening length.
		(a) The screening length $\lambda_c$ in units of $r_0$ in a double-log scale as function of $(B-B_{\rs SIT})/B_{\rs SIT}$. At low magnetic fields $\lambda_c$ follows the power-law dependence shown by the dashed line, $\lambda_c\propto(B-B_{\rs SIT})^{-\nu}$, with $\nu=0.51\pm0.02$.
		At higher magnetic fields the slope increases and the exponent becomes $\nu^{\prime}=2.44\pm 0.04$.
		 (b) Normalized Arrhenius plots of conductances at various fields. The shaded area corresponds to the critical region above $T_{\rs CBKT}$ described by Eqs.\,(\ref{eq:nf}) and (\ref{eq:Eq}). The tails at $T<T_{\rs CBKT}$ demonstrate thermally activated behavior highlighted by the dashed lines. The slopes grow with decreasing magnetic field, the corresponding values $T_0$ of the activation temperature are shown in the inset as black circles.  
		The inset presents a comparison of the activation energies, $T_0$,
		 and
		 $T_{\rs CBKT}\ln(\lambda_c/r_0)$ shown by squares, evidencing the logarithmic dependence of the activation energy on the screening length just as expected for the 2D logarithmic Coulomb interaction between the charges. 
	}
	\label{fig:fig5}
\end{figure}
Let us now discuss the implications of identifying the nature of the Cooper pair insulator as a two-dimensional neutral Coulomb plasma of excessive/deficit Cooper pairs, each carrying the charge $\pm 2e$, and analyze further the parameters of this Cooper pair plasma.
We note first the apparent divergence of $\lambda_c$ upon decreasing $B$.
That $\lambda_c$ depends on $B$ enables its identification as the electrostatic screening length in disordered films, $\lambda_c=\Lambda=\varepsilon d$. Accordingly, its divergence upon decreasing $B$ corresponds to the divergence of the dielectric constant upon approach to the SIT. The value of the magnetic field, $B_{\rs SIT}= 0.015$\,T, where the SIT occurs, is determined by the crossing point of the resistive curves, as seen in Fig.\,\ref{fig:fig2}(b). To analyze the character of the divergence we replot $\lambda_c/r_0$ vs $(B-B_{\rs SIT})/B_{\rs SIT}$ in a double-log scale. Fig.\,\ref{fig:fig5}(a) demonstrates that at the lowest fields $\lambda_c\propto(B-B_{\rs SIT})^{-\nu}$, with $\nu=0.51\pm 0.02$. This reveals the critical character of the divergence of the dielectric constant upon approaching the SIT. 
Our finding is in a full accord with the polarization catastrophe paradigm built on the divergence of $\varepsilon$ as function of the carrier concentration upon approach to the quantum metal-insulator transition, see\,\cite{Paalanen1982} and \cite{Bhatt} for the review. In our experiment, it is $B-B_{\rs SIT}$ that plays the role of the deviation of the carrier concentration from its critical value. 
The peculiarity of our system, that its electrostatic properties are tuned by the magnetic field, may be understood once one realizes that the local suppression of the vortex Bose-condensate proportional to $B-B_{\rs SIT}$ stimulates Landau-Zener tunneling of the Cooper pairs. We would then expect the distance between these Cooper pairs, and hence $\lambda_c$, to scale as $(B-B_{\rs SIT})^{-1/2}$, comparable to what is seen in our experiment. The change of the slope when far from the critical point of the SIT requires further investigation. 

Having determined the screening length and $T_{\rs CBKT}$ as a function of magnetic field, we can now make an independent cross-check on the 2D Coulomb nature of the superinsulator. 
Shown in Fig.\,\ref{fig:fig5}(b) are Arrhenius plots of the normalized conductance vs. $T_{\rs CBKT}/T$ at various magnetic fields highlighting the thermally activated behavior at low temperatures, $T<T_{\rs CBKT}$. Note that Eqs. (\ref{eq:nf}) and (\ref{eq:Eq}) describe conductance only at $T>T_{\rs CBKT}$, shown as a shaded rectangle. The field dependent activation temperatures $T_0$ are presented in the inset to Fig.\,\ref{fig:fig5}(b). 
When the typical size of the unbound pair becomes comparable to $\Lambda$, the interaction ceases to be logarithmic and the conductance is dominated by thermodynamically activated free charges.
Thus, the low-temperature tails in $G(T)$ are expected to be exponential and to depart from the BKT criticality curve. 
Theoretical calculations\,\cite{Fistul,nature} and simulations\,\cite{OrtBV:2015} of thermally activated hopping transport in a 2D insulator with logarithmic Coulomb interactions between the charge carriers yield an activation temperature $T_0\simeq T_{\rs CBKT}\ln(\lambda_c/r_0)$.
In the same inset we present our experimental values of $T_{\rs CBKT}\ln(\lambda_c/r_0)$ at the same fields; these indeed appear remarkably close to the independently determined $T_0$ in accord with the theoretical expectations. This correspondence validates the 2D Coulomb logarithmic interaction between charges at distances not exceeding the screening length. Similarly, exponential low-temperature tails in the resistance were observed in JJA on the superconducting side of the SIT. The tails appeared below the vortex BKT transition temperature where the applied magnetic field introduced the excess unbound vortices\,\cite{Zant}.

We now can resolve the long-standing open question in the study of the SIT: the origin of the giant peak in the magnetoresistance. It arises from the combination of the dielectric constant rapidly decaying with the increase of the magnetic field and the nonmonotonic behavior of $T_{\rs CBKT}$. In order to gain insight into the behavior of the latter, we employ the model of JJA, an array of superconducting granules connected with Josephson links, which is an adequate representation for the critically disordered superconducting film. 
The origin of the nonmonotonic behavior of $T_{\rs CBKT}$ can be explained by recalling that the energy gap of the Cooper pair insulator in JJA, $\Delta_c(B)$, is suppressed by the Josephson coupling $E_{\rs J}$ between the neighboring granules, $\Delta_c(B)=\Delta_c(0)[1-AE_{\rs J}(B)/E_c]$\,\cite{Fistul}, where $E_c$ is the Coulomb energy of a single granule and $A$ is a constant. The Josephson coupling is maximal at zero field and, in the irregular JJA, has the minimum at the frustration factor $f=1/2$\,\cite{Valles}, where $f\equiv BS/\Phi_0$, $S$ is the average area of the JJA elemental cell and $\Phi_0$ is the magnetic flux quantum. Accordingly,
the effective Coulomb energy acquires the \textit{maximum} at $f=1/2$, i.e. the nonmonotonic behavior of $T_{\rs CBKT}$ reflects the nonmonotonic behavior of $E_{\rs J}$ as a function of the magnetic field. This enables us to estimate the parameters of the system as follows.
The observed maximum in $T_{\rs CBKT}$ at $B\approx 4$\,T (\,Fig\,\ref{fig:fig4}(b)) implies that the average area of an elemental cell of our self-induced granular structure, $S\approx 260$\,nm$^2$ and, hence, the linear size of the elemental cell $\sqrt{S}\approx 16$\,nm$\approx 3.5\,\xi$, where $\xi=4.5$\,nm is the superconducting coherence length of the NbTiN film (Table\,II). 
Interestingly, this correlates with the analogous estimates for TiN, where $\sqrt{S}\approx 4\xi$ was observed\,\cite{diode}.    
The described nonmonotonic behavior is accompanied by the overall suppression of the superconducting gap by the increasing magnetic field. The latter eventually would suppress the superconducting gap in Cooper pair droplets and hence $\Delta_c$, resulting in a further drop of the resistance. Then, the Cooper pair insulator ends up as a metal\,\cite{Baturina:2007}.

By comparison to NbTiN, the behaviors complying with the formation of the superinsulating state were observed in other materials at very low temperatures. In TiN films the superinsulator appeared at 40\,mK\,\cite{Kalok:2012}. More recently, the finite temperature zero-conductance state in InO was reported at $T\lesssim 35$\,mK\,\cite{Shahar:2015srep}. The temperature dependence of the conductance in InO was found to follow the so-called Vogel-Fulcher-Tamman dependence, $\sigma\propto\exp[-\mathrm{const}/(T^{\ast}-T)]$\,\cite{Vogel1921,Fulcher1925,Tamman1926}. This, however, can be viewed as a manifestation of the same BKT physics but in a more disordered system\,\cite{SVT:2017}.

To summarize, our findings conclusively establish the finite temperature superinsulating state in NbTiN as the low temperature charge BKT phase of the Cooper pair insulator.  
We demonstrate superinsulating behavior in a new material with a substantially higher transition temperature of nearly 200\,mK, allowing for the first time a detailed characterization of behavior of the system both above and below $T_{\rs CBKT}$ and its evolution in a wide range of magnetic fields.

\subsection{\large{Supplementary Materials}}

\subsection{Fabrication technique}
The Atomic Layer Deposition (ALD) growths were carried out in a custom-made viscous flow ALD reactor in the self limiting regime. A constant flow of ultrahigh-purity nitrogen (UHP, 99.999$\%$, Airgas) at $\sim$ 350 sccm with a pressure of $\sim$ 1.1 Torr was maintained by mass flow controllers. An inert gas purifier (Entegris GateKeeper) was used to further purify the N$_2$ gas by reducing the contamination level of H$_2$, CO, and CO$_2$ to less than 1 ppb and O$_2$ and H$_2$O to less than 100 ppt.
The thermal ALD growth of the AlN/NbTiN multilayer was performed using alternating exposures to the following gaseous reactants with the corresponding timing sequence (exposure-purge) in seconds: AlCl$_3$ (anhydrous, 99.999$\%$, Sigma-Aldrich) (1 - 10), NbCl$_5$ (anhydrous, 99.995$\%$, Sigma-Aldrich) (1 - 10), TiCl$_4$ (99.995$\%$, Sigma-Aldrich) (0.5 - 10) and NH$_3$ (anhydrous, 99.9995$\%$, Sigma-Aldrich) (1.5 - 10).
The intrinsic silicon substrates were initially cleaned in-situ using a 60\,s exposure to O$_3$ repeated 5 times. The AlN buffer layer of thickness 7.5$\pm$0.5\,nm  was deposited at 450$^{\circ}$\,C with 200 ALD cycles. The chamber temperature was then lowered to 350$^{\circ}$\,C to synthesize the NbTiN layers. The growth cycle of the NbTiN is 2$\times$(TiCl$_4$ + NH$_3$) and 1$\times$(NbCl$_5$ + NH$_3$) that was repeated 80, 140, and 256 times with the corresponding total ALD cycles 240, 420 and 768 to produce the different film thicknesses 10\,nm, 15\,nm, and 20\,nm, respectively as measured ex-istu by X-ray reflectivity (XRR). The chemical composition measured by X-ray Photoemission Spectroscopy (XPS) and Rutherford Backscattering Spectroscopy (RBS) show consistently for the AlN layer 5.5$\pm$0.3$\%$ of Cl impurities and a Al/N ratio of 1$\pm$0.05, whereas for the NbTiN films 3$\pm$0.3$\%$ of Cl impurities, a Nb/Ti ratio of 2.3$\pm$0.03 and a (Nb+Ti)/N ratio of 1$\pm$0.03. The material densities measured by RBS and XRR are 2.5$\pm$0.01\,g/cm$^3$ in AlN and 6$\pm$0.05\,g/cm$^3$ in the NbTiN.

\subsection{Analysis of structure and composition of NbTiN films}
The structure of Nb$_{x}$Ti$_{1-x}$N films grown on Si substrates with AlN buffer layers was investigated using the JEOL-4000EX electron microscope operated at 400\,kV, with the point-to-point resolution of 0.16\,nm and the line resolution of 0.1\,nm. 
The Digital Micrograph software (GATAN) was used for the digital processing of High Resolution Transmission Electron Microscopy (HRTEM)
images. 
The interplanar spacings were determined with the accuracy about 0.05\,nm.  
The images displayed in Figs.\,\ref{fig:fig1}(a,b) were calibrated
using the  lattice of the crystalline Si substrate clearly visible in HRTEM image presented in Fig.\,\ref{fig:fig1}(a). In order to determine structure and composition of NbTiN film, we use 
electron diffraction data (Fig.\,\ref{fig:fig1}(c)) which contain characteristic rings and point reflexes. The latter were identified as belonging to the (220) order of the Si substrate with the interplanar spacing $d=1.92$\,{\AA} and then served as a reference scale for determining the interplanar spacing of the NbTiN film.  
\begin{center}
	\begin{table}[h]
		\caption{Lattice Parameters. The
			tabulated parameters for  NbN and TiN, and experimental values for NbTiN.
			$d$ is the interplane distance.} \label{tabular:parametrs}
		\renewcommand{\arraystretch}{1.5}
		\begin{center}
			\begin{tabular}{@{\extracolsep{1cm}}cccc}
				\hline
				\phantom{pha}     & {NbN}        &  {TiN}       &  {NbTiN}     \\
				\hline
				lattice constant & 4.39 [{\AA}] & 4.24 [{\AA}] & 4.33 [{\AA}] \\
				\hline \hline
				\hline plane      & $d$ [{\AA}]  & $d$ [{\AA}]  & $d$ [{\AA}]  \\
				(111)            & 2.54         & 2.45         & 2.51         \\
				(200)            & 2.20          & 2.12         & 2.17         \\
				(220)            & 1.55         & 1.5          & 1.54         \\
				(311)            & 1.32         & 1.28         & 1.31         \\
				(222)            & 1.27         & 1.22         & 1.25         \\
				(420)            & 0.98         & 0.95         & 0.97         \\
				(422)            & 0.90          & 0.87         & 0.89         \\
				\hline \hline
			\end{tabular}
		\end{center}
	\end{table}
\end{center}
The detailed analysis of the electron-diffraction data reveals that our NbTiN have the same rock-salt crystalline structure as both NbN and TiN.
We find that the lattice constant of the NbTiN film is $a = 4.33$ {\AA}.
The values of the interplanar spacings corresponding to the different planes are derived from the positions of the ring brightness maxima shown in Fig.\,\ref{fig:fig1}(d) and are presented in 
Table I. 
The AlN buffer layer has hexagonal lattice 63\,mc; the (100) order of the buffer layer has $d = 2.7$\,{\AA}. 
The composition of the NbTiN films is found using the
Vegard's law, 
\begin{equation}
	a_{\textrm{NbTiN}}= x\cdot a_{\textrm{NbN}}+(1-x)\cdot
	a_{\textrm{TiN}},
	\label{Vegards}
\end{equation}
where  $a$ is the lattice constant and $0 \leq x \leq 1$.
Comparison the tabulated data on NbN and TiN (see Table\,I) yields that NbTiN is a solid solution of NbN~\cite{Handbook, NbN} and TiN~\cite{Handbook,TiN} with $x = 0.7\pm0.02$.

\subsection{Samples and Measurements}
In order to carry out transport measurements, NbTiN films were patterned using photolithography and plasma etching into 10-contact resistivity bars 50\,$\mu$m wide and with 100, 250, and 100\,$\mu$m separation between the voltage probes (Fig.\,\ref{fig:fig1_SI}). The chosen design allows for Hall effect measurements and both two-probe and four-probe resistivity measurements.
\begin{figure}[h]
	\begin{center}
		\includegraphics[width=0.35\textwidth]{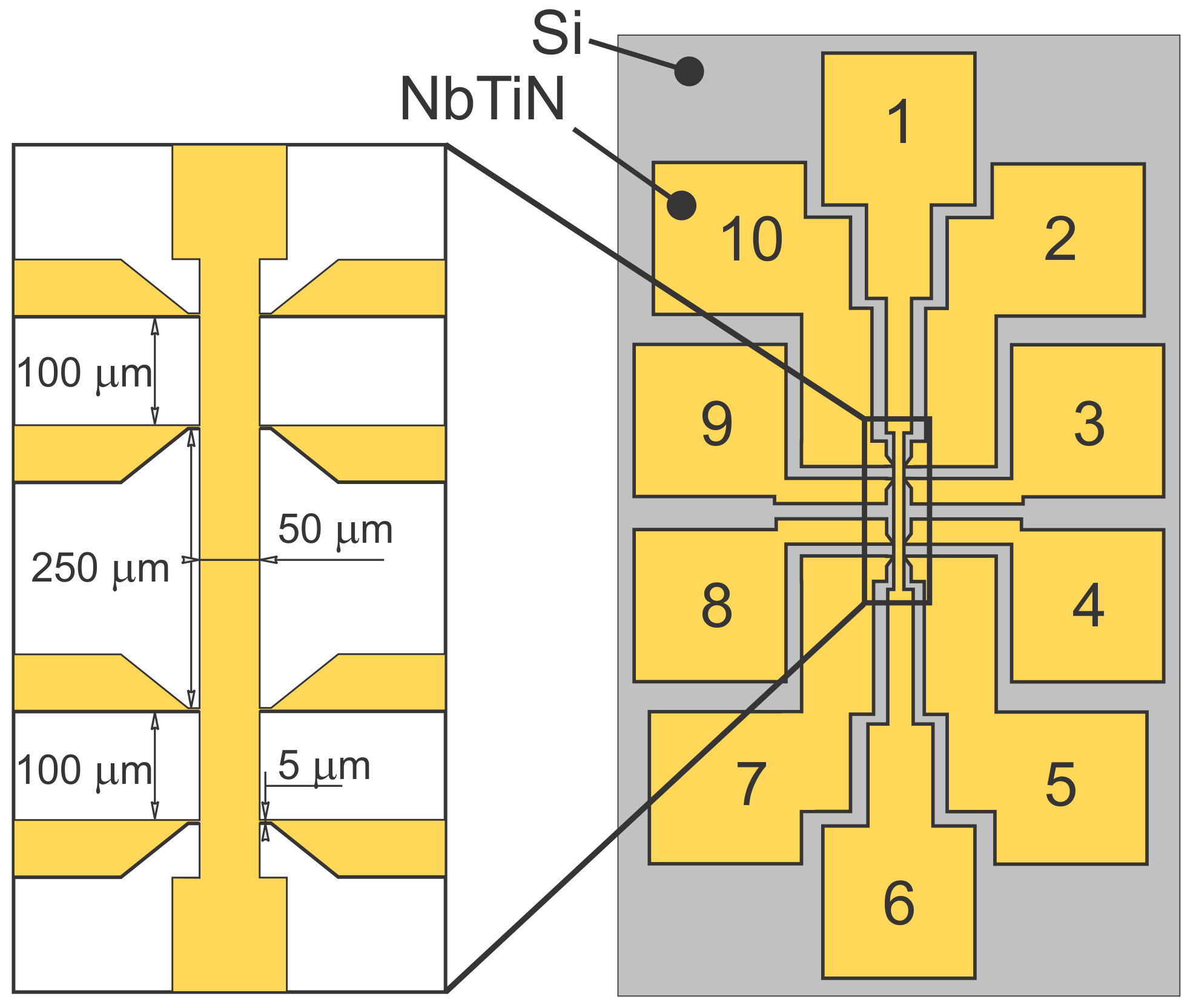}
	\end{center} 
	\caption{The sketch of the samples.}
	\label{fig:fig1_SI}
\end{figure}
Measurements of the temperature and magnetic field dependences of the resistance were carried out in helium dilution refrigerators. The magnetic field was always perpendicular to the plane of the sample. In the low resistance range, the standard four-probe constant-current measurements were performed at 10\,nA and 3\,Hz. In the high resistance range, the two-probe constant-voltage technique was used instead, with a 100 $\mu$V/3 Hz\,probe. The excitations in both cases were verified to be in the linear response regime. Both sets of measurements were performed using SR830 lock-in amplifiers. The high-resistivity measurements also employed an SR570 low-noise current preamplifier.
The resistance per square in the two-probe geometry was determined by matching the two-probe and four probe measurements at high temperature ($T \sim 10$~K).
\begin{center}
	\begin{table}[h]
	\caption{Parameters of NbTiN films:  $d$ is the film thickness, $T_c$ is the superconducting critical temperature defined by the inflection point of $R(T)$, $R_{300}$ is the room temperature resistance, $B_{c2}(0)$ is the upper critical field, $D$ is the diffusion constant, $\xi_{d}(0)$ is the superconducting coherence length.}
		\label{tabular:tr_parametrs}
		\renewcommand{\arraystretch}{1.3}
		\begin{center}
			\begin{tabular}{@{\extracolsep{0.2cm}}cccccc}
				\hline \hline
				$d$ [nm] & $T_c$ [K] & $R_{300}$ & $B_{c2}(0)$\,[T] & $D$ [cm$^{2}$/c] & $\xi_{d}(0)$ [nm]\\
				\hline
				10      & $0.33$    &  1900     &  ---     & ---              &   ---             \\
				15      & $3.35$    &  940      &  10.5    & 0.24             &   4.65          \\
				20      & $4.27$    &  390      &  12      & 0.27             &   4.35          \\
				\hline \hline
			\end{tabular}
		\end{center}
	\end{table}
\end{center}
The measurement of the Hall effect in 10\,nm thick film yields the carrier density $n = 4.7\cdot10^{21}$\,cm$^{-3}$. 
The upper critical field $B_{c2}(0)$ for films 15 and 20\,nm was estimated as $B_c/1.05$, where $B_c$ 
is the crossing point\,\cite{Baturina2005,Finn}.
The crossing point for the 15\,nm thick film is shown in the Fig.\,2(c). 
The diffusion coefficient, $D$, and the superconducting coherence length, $\xi_d(0)$, are given by
\begin{equation}\label{DiffCoeff}
	D = \frac{\pi k_B T_c}{2 \gamma e B_{c2}(0)},
\end{equation}

\begin{equation}\label{xi_d}
	\xi_d(0) = 0.85\sqrt{\xi_0 l} = 0.85\sqrt{\frac{3}{2\pi}\frac{\hbar}{eB_{c2}(0)}}\,,
\end{equation}
where $k_{\rs B}$ is Boltzmann constant, $\gamma$ is  Euler's constant, and $\gamma \approx 1.781$.

\subsection{Acknowledgments}
This work was supported by the Ministry of
Education and Science of the Russian Federation, by the Grant of the President
RF, project No MK-4628.2016.2 (AYuM, SVP, MVB, and AKG). The high
resolution electron microscopy was performed with support of RSF, project
No 14-22-00143. The work at Caltech was supported by National Science
Foundation Grant No. DMR-1606858 (DMS and TFR). The work at Argonne
was supported by the U.S. Department of Energy, Office of Science, Materials
Sciences and Engineering Division (VMV, TP, and visits of TIB). TIB also
acknowledges support by the Alexander von Humboldt Foundation and from
the Consejería de Educación, Cultura y Deporte (Comunidad de Madrid)
through the talent attraction program, Ref. 2016-T3/IND-1839. AYuM and
TIB were also supported from the Argonne-University of Chicago collaborative
seed grant.

\end{document}